\documentclass[twocolumn,showpacs,prl,aps,tightenlines,superscriptaddress]{revtex4}

\usepackage{graphics,graphicx}

\begin{document}

\title{Supermassive recoil velocities
for binary black-hole mergers with antialigned spins}

\author{Jos\'e A. Gonz\'alez, Mark Hannam, Ulrich Sperhake, Bernd Br{\"u}gmann, Sascha Husa}
\affiliation{Theoretical Physics Institute, University of Jena, 07743 Jena, Germany}

\begin{abstract}
Recent calculations of the recoil velocity in binary black hole mergers
have found the kick velocity to be of the order of a few hundred
km/s in the case of non-spinning binaries and about $500\,$km/s in the
case of spinning configurations, and have lead to predictions of a maximum kick 
of up to $1300\,$km/s. We test these predictions and demonstrate that kick velocities 
of at least $2500\,$km/s are possible for equal-mass binaries with
anti-aligned spins in the orbital plane. Kicks of that magnitude are likely to
have significant repercussions for models of black-hole formation, the
population of intergalactic black holes and the structure of host galaxies.
\end{abstract}


\maketitle

\paragraph{Introduction.---}

A well known phenomenon of general relativity is the recoil or rocket
effect due to the emission of anisotropic gravitational radiation
\cite{Bonnor1961, Peres1962, Bekenstein1973}. The loss of linear momentum
radiated away in the form of gravitational waves imparts a recoil or kick
on the remaining system which then moves relative to its original centre-of-mass 
frame. This effect is particularly pronounced for the inspiral
and merger of two compact objects and thus may have dramatic
consequences for the merger of massive black holes residing at the
centers of galaxies when their hosts undergo merger.

Massive black holes with masses $10^5$ to $10^{9.5}\,M_{\odot}$
are not only known to exist at the centre of many galaxies, but also to have
a substantial impact on the structure and formation of their host galaxies,
as is demonstrated by the correlation of the black-hole mass
with the bulge mass, luminosity and velocity dispersion
\cite{
Magorrian1998etal, 
Gebhardt2000etal,
Merritt2001, McLure2002}.
The merger of galaxies is thus likely to be accompanied by an inspiral
of the central black holes. 
In order to assess
the impact of the resulting black-hole recoil on questions
such as the bulge structure, the formation history of massive
black holes and interstellar and intergalactic populations
of black holes, it is vital to have a good understanding of the
kick magnitude and, in particular, the maximum possible kick velocities.

The first study of the recoil effect for inspiralling binaries was performed
by Fitchett \cite{Fitchett1983} in the framework of non-spinning point particles
subject to Newtonian gravity. The resulting recoil calculated using the
lowest order multipoles is given in his Eq.\,(3.21) and predicts
kick velocities of the order of $1000\,$km/s that exceed
 the escape velocities from massive galaxies \cite{Redmount1989}. The
problem was reinvestigated using the particle
approximation \cite{Fitchett1984, Nakamura1983, Favata2004, Lousto2004},
post-Newtonian methods \cite{Wiseman1992, Kidder1995, Blanchet2005,
Damour2006}, the Close-Limit Approximation \cite{Andrade1997,
Sopuerta2006a, Sopuerta2007} as well as numerical simulations
\cite{Anninos1998, Campanelli2005}. The picture that emerged from
these studies is that the recoil from unequal-mass, non-spinning
binaries is unlikely to exceed a few hundred km/s.

Recent breakthroughs in the numerical simulation of black-hole binaries
\cite{Campanelli2006, Baker2006, Pretorius2005} 
have enabled investigations
of the recoil problem without any restrictive approximations
other than the numerical differencing of the Einstein equations. First
studies addressed the recoil from non-spinning binaries and confirmed the
relatively small magnitude of the kick velocities for mass ratios
in the range 1:1 to 1:2 \cite{Herrmann2006, Baker2006betal}. In a
previous publication \cite{Gonzalez2006}, we have presented the most
comprehensive study of this problem and found a maximum kick of
$175.7\pm11\,$km/s for a mass ratio $\eta=m_1m_2/(m_1+m_2)^2=0.195\pm0.005$.

More recent numerical studies have shown, however, that as expected 
significantly larger kicks are realized if one allows at least one black hole to spin.
 Simulations of
equal-mass binaries with spins orthogonal to the orbital plane 
predict kick velocities of $475$ and $440\,$km/s, respectively, in the
limit of extreme Kerr black holes \cite{Herrmann2007,Koppitz2007etal}. Kicks of
tens of km$/$s have been obtained from head-on collisions of spinning black
holes \cite{Choi:2007eu}. These results are consistent with the
effective-one-body post-Newtonian (PN) study 
in \cite{Schnittman:2007sn}. 
In addition, Campanelli, {\it et. al.} \cite{Campanelli2007v1}
obtain $v_{\rm kick}=454\,$km/s in the case of a non-spinning black
hole orbiting a spinning counterpart of twice the mass with the spin
oriented at $-45^{\circ}$ relative to the orbital plane. They further estimate
that an equal-mass binary with spins aligned in the orbital plane could 
produce a maximum kick of around $1300\,$km/s based on a PN calculation
by Kidder \cite{Kidder1995}. 

In this letter we investigate the scenario suggested
by  Campanelli, {\it et. al.} \cite{Campanelli2007v1,Kidder1995}
and find that a kick of $2500\,$km/s is possible --- larger than the escape
velocity of about $2000\,$km/s of giant elliptical galaxies.

\paragraph{Numerical framework.---}

The numerical simulations presented in this work were performed independently
with the {\sc Bam} and the {\sc Lean} codes. These codes employ the
moving-puncture method \cite{Campanelli2006,Baker2006},
and are described in detail
in \cite{Bruegmann2006aetal,Sperhake2006}. We note, however,
the following modifications
of the Cactus \cite{Cactusweb} and Carpet \cite{Schnetter2004} based
{\sc Lean} code relative to the version presented in \cite{Sperhake2006}:
time evolution of the {\sc TwoPuncture} initial data \cite{Ansorg2004}
is performed using the fourth-order accurate
Runge-Kutta method, the variable $\phi$ 
has been replaced by the new variable $\chi=e^{-\phi}$ and the code
uses fifth-order prolongation in space. 
The two codes represent independent implementations of similar
techniques, thereby allowing important cross-validation.

The gravitational waves emitted by the binary are extracted by computing the
Newman-Penrose scalar $\Psi_4$ over spheres of constant radius at different
distances. The notation and equations used for this procedure are explained in 
detail in section III.A of \cite{Bruegmann2006aetal}. 
In order to estimate the recoil velocity of the system, we compute 
the total linear momentum radiated during the simulation directly from $\Psi_4$ 
using equation (1) in \cite{Gonzalez2006}. The pulse of spurious radiation from 
the initial data, which would lead to an over-estimation of the final kick by about 5\%, is
not included in this calculation. We have neglected any
recoil that may have accumulated during the earlier inspiral of the black holes; however,
for the final kick values we obtain, this introduces an uncertainty of less than 1\% in our
results (as obtained from PN estimates in the nonspinning case \cite{Blanchet2005}), 
which is far less than the uncertainty due to other errors.

We consider configurations with spins of equal size but opposite direction lying 
in the orbital plane. 
The initial parameters of Model I (MI) were chosen without particular regard to obtain a 
quasi-circular configuration. On the other hand, the initial momenta used
for Model II (MII) were computed using formula $(4.7)$ from~\cite{Kidder1995} to
generate a quasi-circular orbit. The parameters and the values of the kick 
obtained for the two models are summarized in Table~\ref{tbl:1}. 
The initial configurations, consisting of two equal-mass black holes
with spins perpendicular to the orbital angular momentum, are chosen to
maximize Eq. (3.31b) in~\cite{Kidder1995} as suggested in~\cite{Campanelli2007v1}.
Each black hole has a total mass of $m \sim 0.5$ and we consider spin parameters of
the order of $a/m \sim 0.8$, which is below 
the recent estimates presented in \cite{McClintock2006etal}.

\begin{table}
\begin{ruledtabular}
\begin{tabular}{l|r|r|r|r|r}
Run       & $X$         & $P_y$        & $m_i$   & $S_x$     & $v_{\rm kick} (km/s)$ \\
\hline
{\sc MI}  & $\pm 3.257$ & $\pm 0.133$  & $0.363$ & $\pm 0.2$ & $2450\pm250$  \\
{\sc MII} & $\pm 4.0$   & $\pm 0.1125$ & $0.287$ & $\pm 0.2$ & $2650$  \\
\end{tabular}
\end{ruledtabular}
\caption{Initial puncture parameters and final kick velocity. \label{tbl:1}}
\end{table}

\paragraph{Results.---}

\begin{figure}[h]
\includegraphics[width=6.8cm,height=9cm,angle=-90]{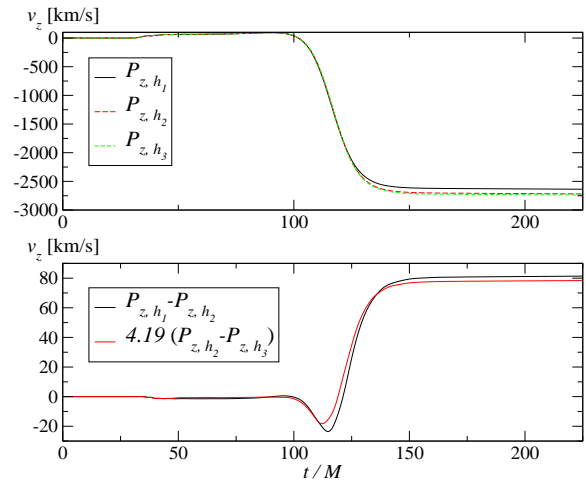}
\caption{\label{fig:Pz_conv} Upper panel: the $z$-component of the
recoil for model MI as a function of time for grid resolutions
$h_1=1/36$, $h_2=1/44$ and $h_3=1/48$. Lower panel: the differences
scaled for fourth-order convergence.}
\end{figure}

We performed 
evolutions of model MI with the {\sc Lean} code using
different resolutions and computed the radiated linear 
momentum at different extraction radii. 
In Figure~\ref{fig:Pz_conv} we present the recoil speed $v_z = - P_z/m$ of the
final black hole as a function of time for the resolutions $1/36$, $1/44$ and
$1/48$. In this context we note that the $x$ and $y$ components of $v$ are
smaller than $1\,$km/s, so that the $z$-component is practically identical
with the total recoil. The bottom panel of the figure demonstrates
fourth-order convergence of the recoil, and we estimate the uncertainty as
$43\,$km/s or $1.5\,\%$.

\begin{figure}[h]
\includegraphics[width=6.8cm,height=9cm,angle=-90]{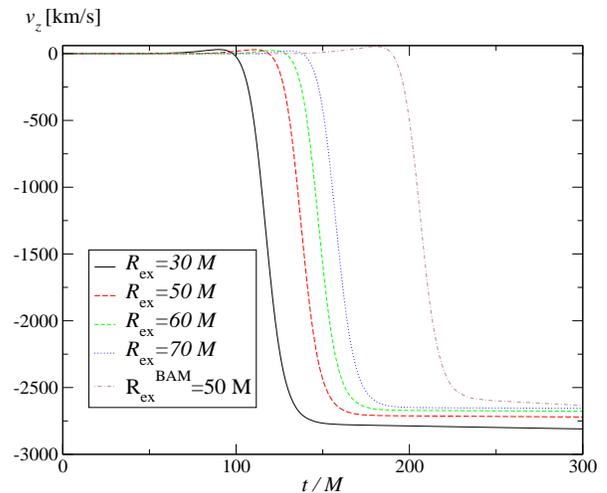}
\caption{\label{fig: Pz_Rex} Recoil for model MI as a function of time
  for different extraction radii. Extrapolation to $r\to\infty$ yields a
  magnitude of $2450\,$km/s with an error estimate of
  $4.5\,\%$. Also shown is the kick
  from model MII using {\sc Bam}.} 
\end{figure}

The dependence of the recoil velocity on the extraction radius is
presented in Figure~\ref{fig: Pz_Rex}. We find the resulting error to
be reasonably well approximated by a $1/r_{\rm ex}$ falloff and thus obtain
uncertainties due to finite extraction radii of $120\,$km/s or
$4.5\,\%$. Using a conservative error estimate, the magnitude of the final
kick is $2450 \pm 250\,$km/s.

In order to obtain an independent estimate of the kick magnitude for this
type of spin-configuration, we evolve model MII using the {\sc Bam} code. This
model starts at larger initial separation and represents a quasi-circular
configuration. These simulations thus enable us to assess the uncertainties
arising from finite black hole separation and deviations from
circularity.

\begin{figure}[h]
\includegraphics[width=6cm,height=6cm]{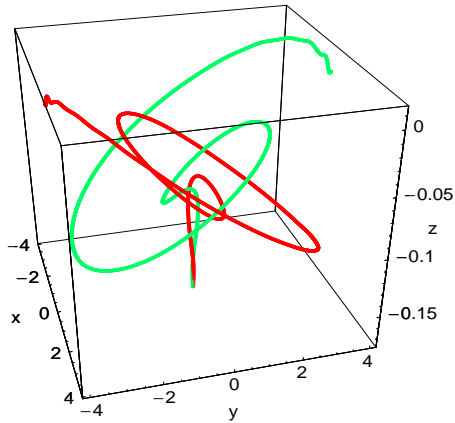}
\caption{\label{fig:fig3} Coordinate positions of the black-hole punctures for
  model MII up to $t = 180$. The black holes move out of the original plane
  and after merger the final black hole receives a kick in the negative $z$-direction.}
\end{figure}

The trajectories of the black holes are shown for model MII
in Figure~\ref{fig:fig3}
and demonstrate how the black holes move out of the initial orbital
plane and acquire substantial momentum in the $z$-direction after the merger.

In summary, our simulations predict recoil velocities of $2500\,$km/s with a
conservative error estimate of $10\,\%$ or $250\,$km/s. 
This value can be compared with the escape velocities from
dwarf elliptical and spheroidal galaxies ($\lesssim300$km$/$s)  and 
from giant elliptical galaxies ($\lesssim2000$km$/$s)~\cite{Merritt2004}. 

Our prediction is larger than any previous
numerical result by a factor of five, but other numerical studies
considered quite different initial configurations. Since the estimate
of $1000\,$km/s \cite{Campanelli2007v1} for the spins studied here
involves a post-Newtonian argument, it is perhaps not surprising that
a fully relativistic simulation obtains a different result.
This result is very intriguing, 
but we point out that so far only the two data points presented here
are available. It will be important to perform studies at higher
numerical resolution, and to study neighboring data sets, for example
to vary the initial separation, spin configuration and mass ratio of
the black holes.

\paragraph{Astrophysical relevance.---}

Before we discuss in more detail astrophysical implications of our
findings, we emphasize one additional important point: while our results
demonstrate that kicks as large as $2,500\,$km/s are possible provided the
inspiralling black holes have appropriate spin alignment and magnitude,
it does not enable us to make any statements on the probability
that these kicks are realized in typical astrophysical merger scenarios.
A conclusive assessment of the astrophysical implications therefore
requires systematic parameter studies as mentioned in the previous section.

Recoil velocities close to or exceeding
the escape velocities of the black hole's
host give rise to a population of black holes away from the galactic nuclei
\cite{Volonteri2003}. 
In contrast to the non-spinning case, the recoil velocities calculated
from our simulations exceed the escape velocities of even massive galaxies
and thus imply larger populations of wandering black holes.
Observational consequences of wandering black holes have been discussed
in detail in \citet{Volonteri2005}.

Black-hole recoil has also been found to give rise to a
core formation in the
central density profile in the host stellar bulge. \citet{BoylanKolchin2004}
find this effect to be most pronounced for recoil
velocities just below the escape velocity. 

Libeskind, {\it et. al} \cite{Libeskind2006} have further found that
gravitational recoil manifests itself as a scatter
of the relation between black hole and bulge mass.
They find this effect to be sensitive to the stability of the disc under
galaxy merger. Their simulations indicate a constraint of
$v_{\rm kick} \lesssim 500\,$km/s. Merritt, {\it et. al.} \cite{Merritt2006etal}
find constraints of a similar magnitude from narrow emission-line analysis
of quasar spectra.

Ejection of black holes affects the rate of binary black hole inspirals
in globular clusters and thus the predicted event rates for gravitational
wave observatories LIGO, GEO600, TAMA and VIRGO. The latest kick
estimates for non-spinning binaries have been taken into account by 
O'Leary, {\it et. al.} \cite{OLeary2007}. It will be of
interest to estimate the impact of spins on such simulations.

The formation history of massive black holes at $z \sim 6$ is often described in
the context of hierarchical 
structure formation via accretion and merger of black holes residing in
dark matter halos (see, e.\,g.\,\cite{Volonteri2003}).
The ejection of black holes from their hosts via
gravitational recoil puts constraints on the maximum redshift at which
the progenitor seed holes might have started merging
and might necessitate accretion growth above the Eddington limit
\cite{Haiman2004, Merritt2004} and may also lead to a population
of intermediate mass black holes \cite{Volonteri2005}. 
Black hole mergers would not occur if seed black holes are rare
at high redshifts, so that black-hole binaries would not commonly
form as a consequence of dark matter halo mergers \cite{Madau2004}.

Finally, black hole recoil has been suggested to manifest itself directly
in observations of off-centre radio-loud active galactic nuclei
\cite{Madau2004} and off-nuclear ultraluminous X-ray sources in nearby
galaxies \cite{Colbert1999}.

Astrophysical studies of the consequences of kick velocities on these
scenarios have so far commonly assumed recoil velocities significantly below
the values obtained in this work. It will be interesting to see 
how these consequences are affected by our results. It will also be important
to estimate the kick magnitudes in a far wider volume of parameter space.

{\em Note added after preparation of this manuscript:} After the first
publication on gr-qc of the results reported here, there have been two
independent computations of similarly large kicks, first reported in
\cite{Herrmann2007b} and in an updated version of \cite{Campanelli2007v1}. Since
our work is based on the original version of \cite{Campanelli2007v1}, it is
relevant to note that the authors have now added new numerical data to their
preprint for a configuration showing a kick of
$1830\,$km/s. They have also modified their argument for the extrapolation
from post-Newtonian results (see also \cite{Baker2007}) now obtaining an upper
limit of $4000\,$km/s \cite{Campanelli2007}. This is consistent with the
results we report here and supports the main point we make, namely that
kicks much larger than previously expected are indeed realized in numerical
simulations with potentially great significance for astrophysics.

\acknowledgments
This work was supported in part by
DFG grant SFB/Transregio~7 ``Gravitational Wave Astronomy''.
We thank the DEISA Consortium (co-funded by the EU, FP6 project
508830), for support within the DEISA Extreme Computing Initiative
(www.deisa.org); computations were performed at LRZ Munich and HLRS, Stuttgart.
J.G. and U.S. acknowledge support from the ILIAS Sixth Framework Programme.

\bibliography{uli}

\end{document}